\documentclass[aps,amsmath,amssymb,nofootinbib]{revtex4}
\usepackage{epsf}
\usepackage{graphicx}
\usepackage{subfigure}
\begin{document}
\title{Closing the Wedge with 300 fb$^{-1}$ and 3000 fb$^{-1}$ at the LHC: A Snowmass White Paper}
\author{Ian~M.~Lewis}

\affiliation{
Department of Physics,
Brookhaven National Laboratory\\
Upton, N.Y.~ 11973, USA}

\date{\today}

\begin{abstract}
The discovery of a Higgs boson at the LHC begins the era of directly measuring the mechanism of electroweak symmetry breaking (EWSB).  Searching for extensions of the Standard Model (SM) EWSB sector at the LHC is of vital importance.  An important extension of the SM with an extended EWSB sector is the Minimal Supersymmetric Standard Model (MSSM).  In this white paper, we extend current ATLAS and CMS bounds on direct searches for the heavy MSSM neutral Higgs bosons to 300 fb$^{-1}$ and 3000 fb$^{-1}$ of data at the LHC.  In particular we focus on the $\tau^+\tau^-$ channel and the pseudoscalar decay to light Higgs boson and Z, with additional discussion on how to further close the wedge.
\end{abstract}

\maketitle
Now that a Higgs boson has been discovered~\cite{Aad:2012tfa}, the era of directly probing the mechanism of electroweak symmetry breaking (EWSB) has begun.  Many beyond the Standard Model (SM)~\cite{Glashow:1961tr} theories contain expansions of the EWSB sector.  Searching for these extensions is important for the full understanding of EWSB.  One such popular theory is Supersymmetry (SUSY). For phenomenological studies at the LHC, it is useful to focus on the Minimal Supersymmetric Standard Model (MSSM)~\cite{Nilles:1983ge}, which is the minimal low-energy realization of a supersymmetric model that contains the SM.

In the MSSM, two Higgs doublets are needed for anomaly cancellation.  One doublet gives mass to up-type fermions and the other to down-type fermions, the so-called type-II two Higgs doublet model.  As a result, after EWSB and three Goldstone Bosons are eaten by $W^\pm$ and $Z$, there are five Higgs bosons.  Assuming no CP violation in the Higgs sector, the MSSM Higgs bosons are a light neutral CP-even Higgs, $h$, a heavy neutral CP-even Higgs, $H$, a neutral CP-odd Higgs, $A$, and two charged Higgs, $H^\pm$.  At tree level, the Higgs sector depends on two parameters, typically chosen to be the mass of the pseudoscalar, $M_A$, and the ratio of the vevs of the two Higgs fields, $\tan\beta=v_1/v_2$.

As is well-known, at tree level the mass of the lightest MSSM Higgs boson, $M_h$, is bounded from above by the $Z$-mass.  However, loop corrections~\cite{Okada:1990vk} can raise $M_h$ to the observed value\footnote{In this study we assume that the discovered boson is the light neutral CP-even Higgs boson.  For recent studies of the alternative, where the observed state is the heavy neutral CP-even Higgs, please see the references in~\cite{Christensen:2012ei}.} of $M_h=125.5^{+0.5}_{-0.6}$~GeV as measured by ATLAS~\cite{ATLAS:2013mma} and $M_h=125.7\pm0.4$~GeV as measured by CMS~\cite{CMS:yva}.  The leading SUSY corrections to $M_h$ arise from stop loops, with the leading component being
\begin{eqnarray}
\varepsilon = \frac{3\overline{m}_t^4}{2\pi^2 v^2 \sin^2\beta}\left[\log\frac{M^2_S}{\overline{m}_t^2}+\frac{X^2_t}{M^2_S}\left(1-\frac{X^2_t}{12M^2_S}\right)\right],
\end{eqnarray}
where $\overline{m}_t$ is the running $\overline{\text{MS}}$ top mass, $X_t=A_t-\mu\cot\beta$ is the stop mixing parameter, $M_S=\sqrt{m_{\tilde{t}_1}m_{\tilde{t}_2}}$ is the geometric mean of the two stop squark masses, and $v^2=v^2_1+v^2_2=(246~{\rm GeV})^2$. 

\begin{figure*}[tb]
\centering
\subfigure[]{\includegraphics[width=0.3\textwidth,clip]{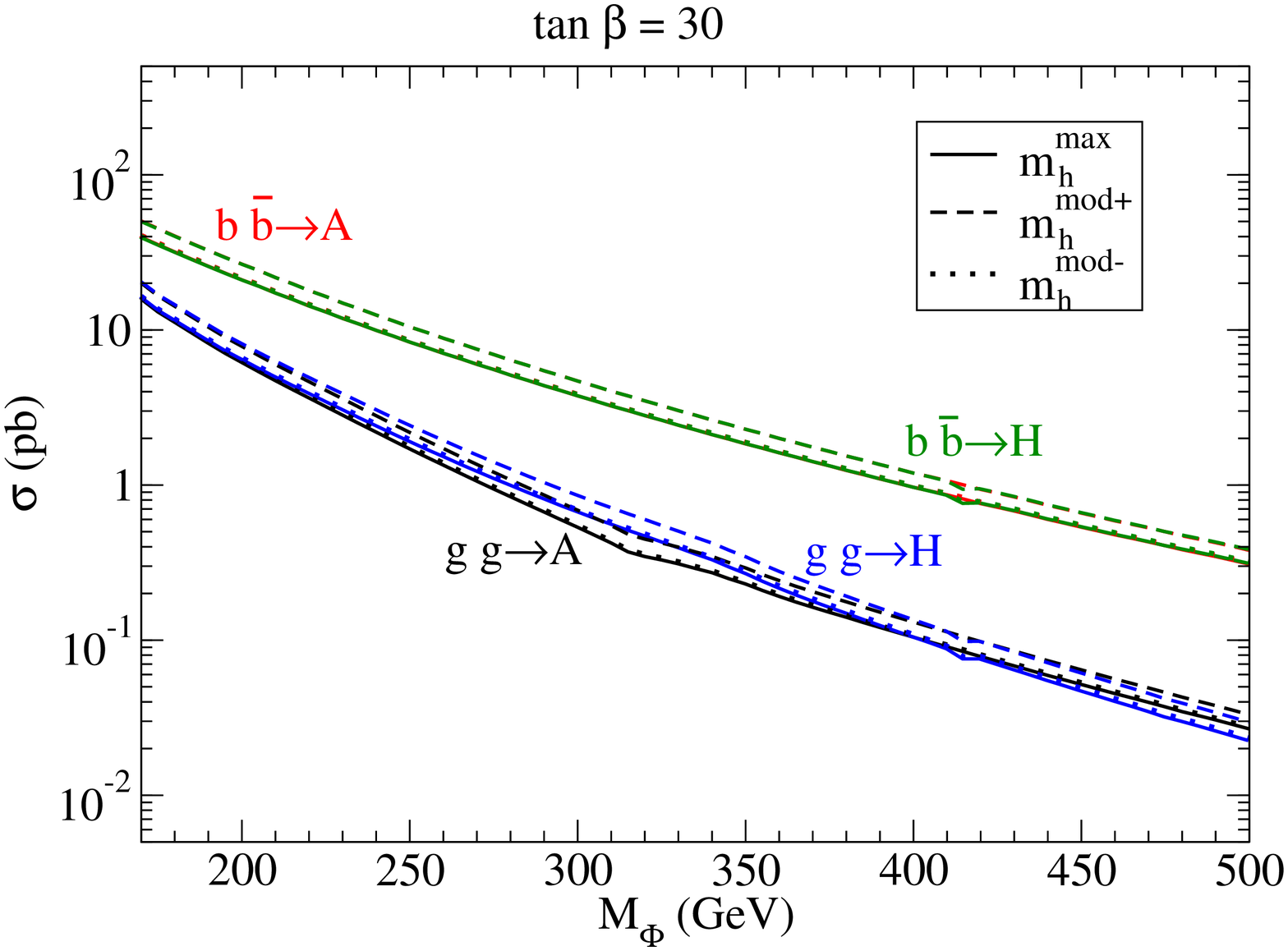}}
\subfigure[]{\includegraphics[width=0.3\textwidth,clip]{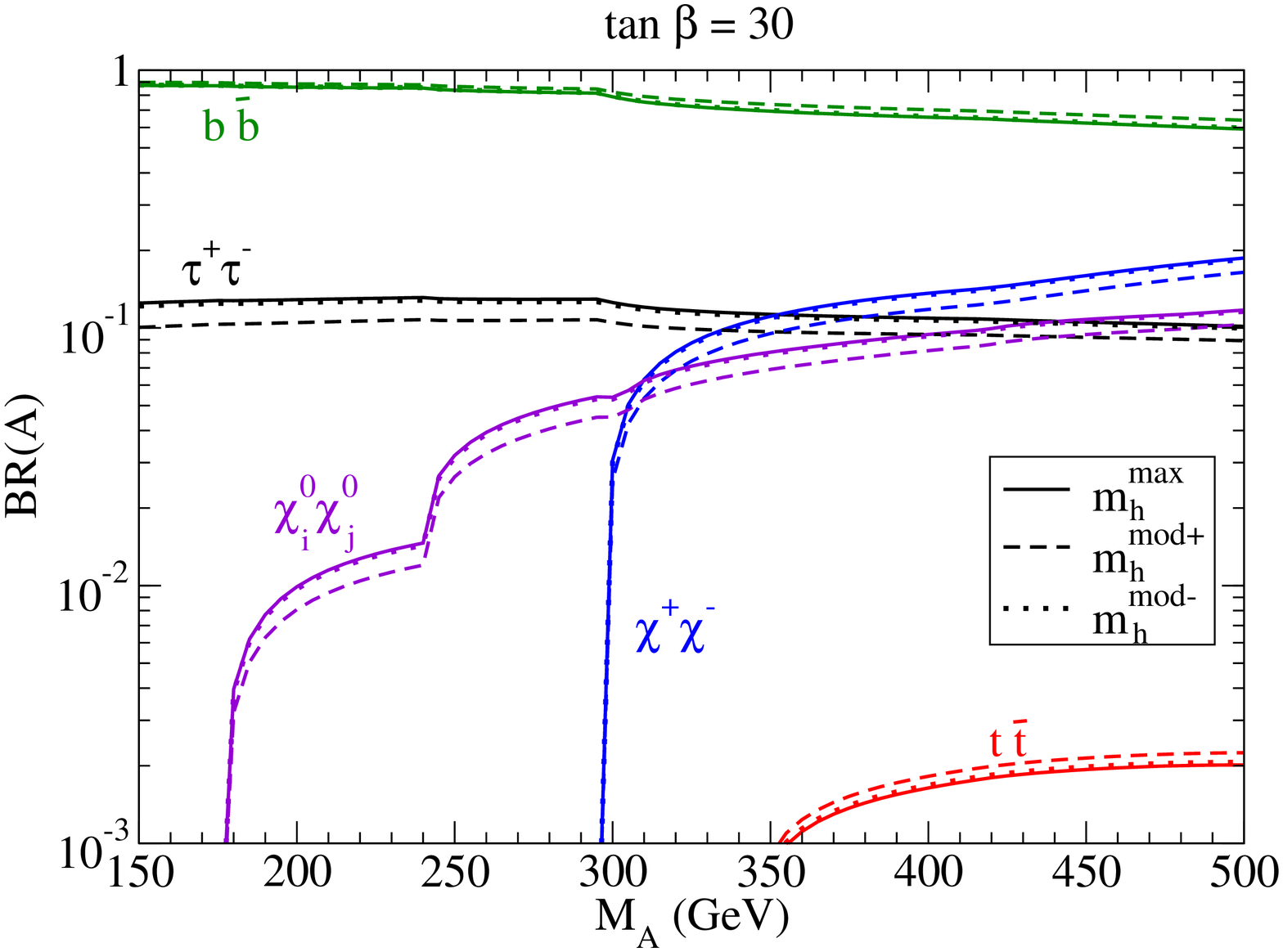}}
\subfigure[]{\includegraphics[width=0.3\textwidth,clip]{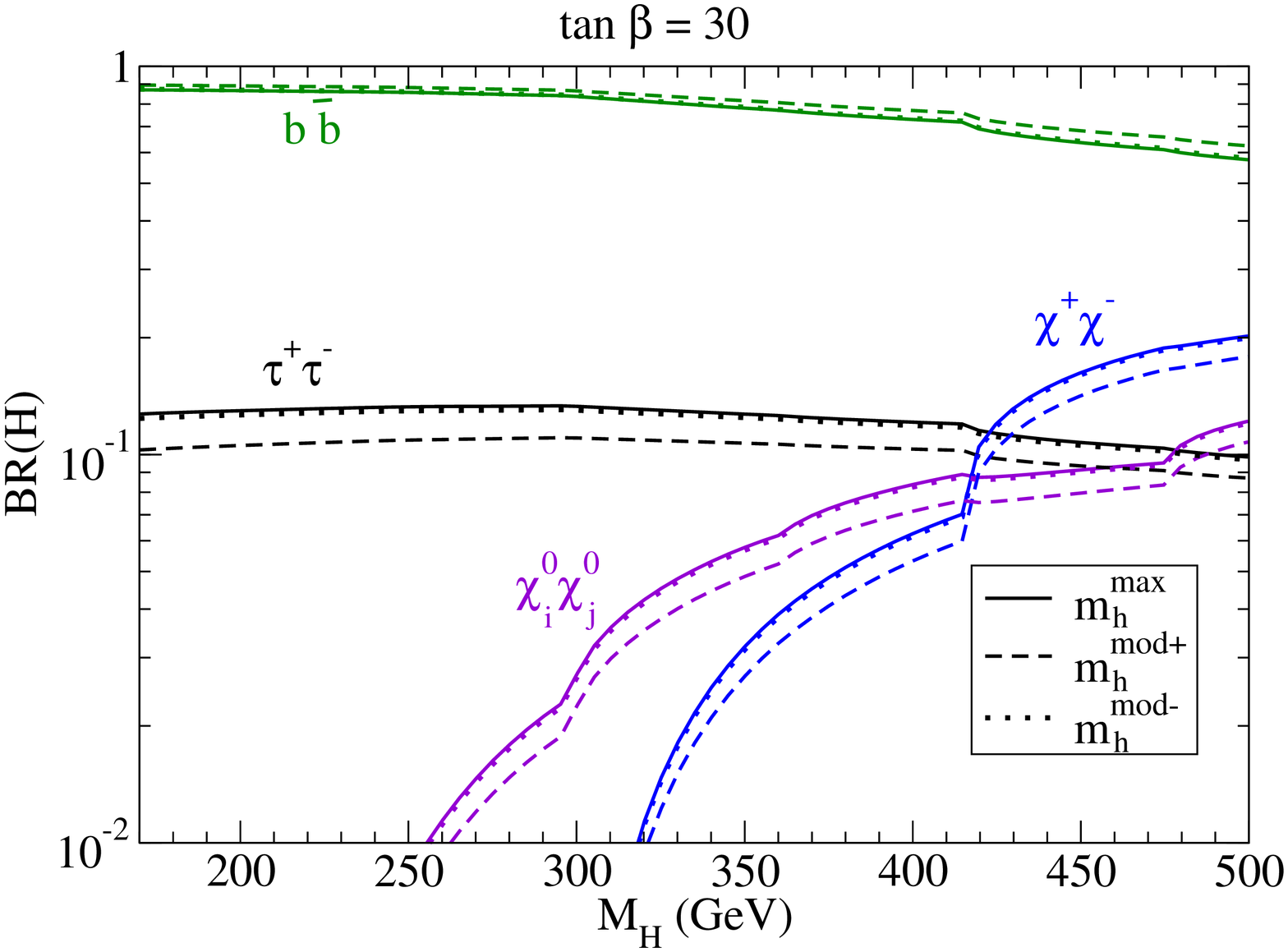}}\\
\subfigure[]{\includegraphics[width=0.3\textwidth,clip]{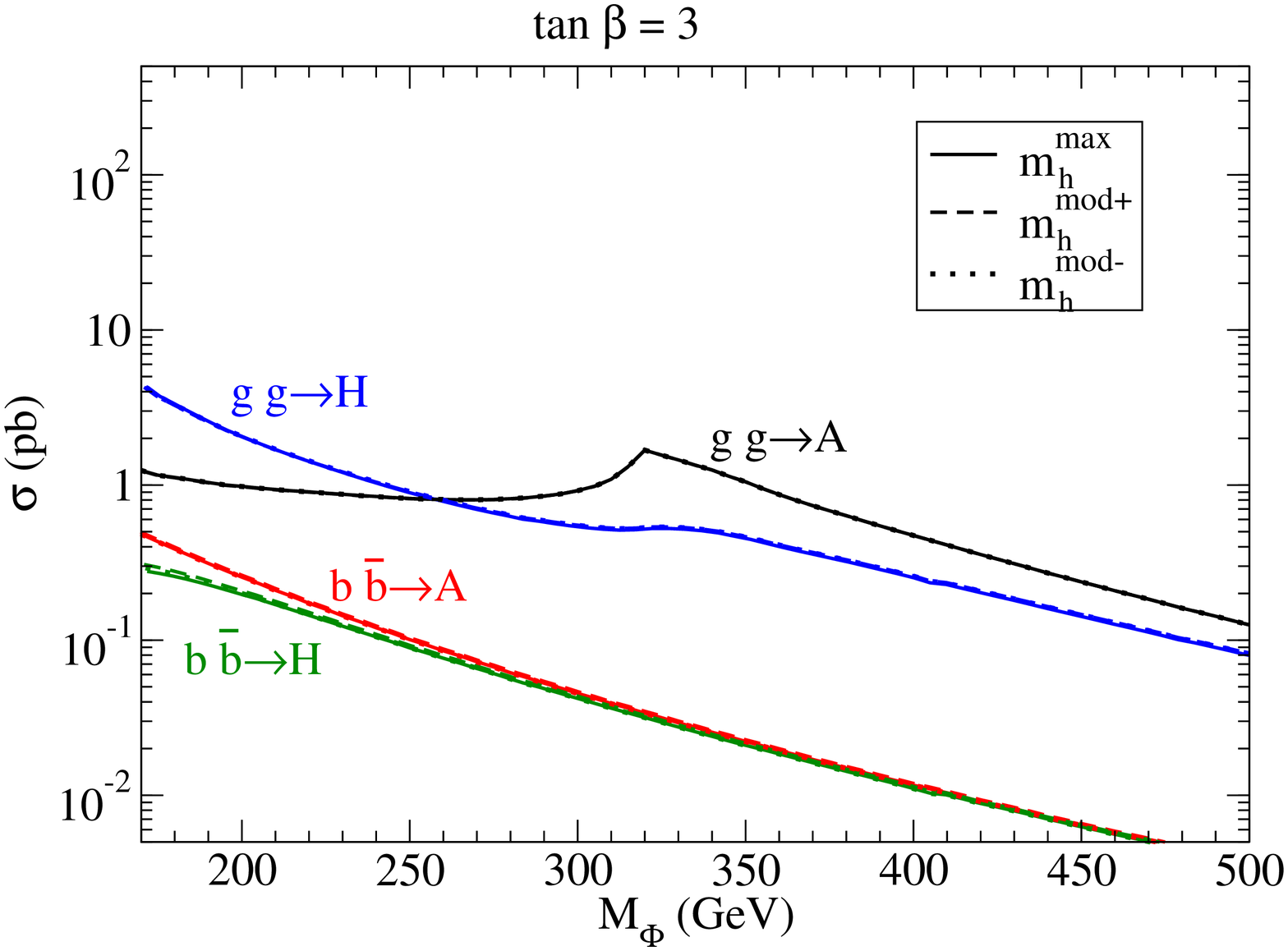}}
\subfigure[]{\includegraphics[width=0.3\textwidth,clip]{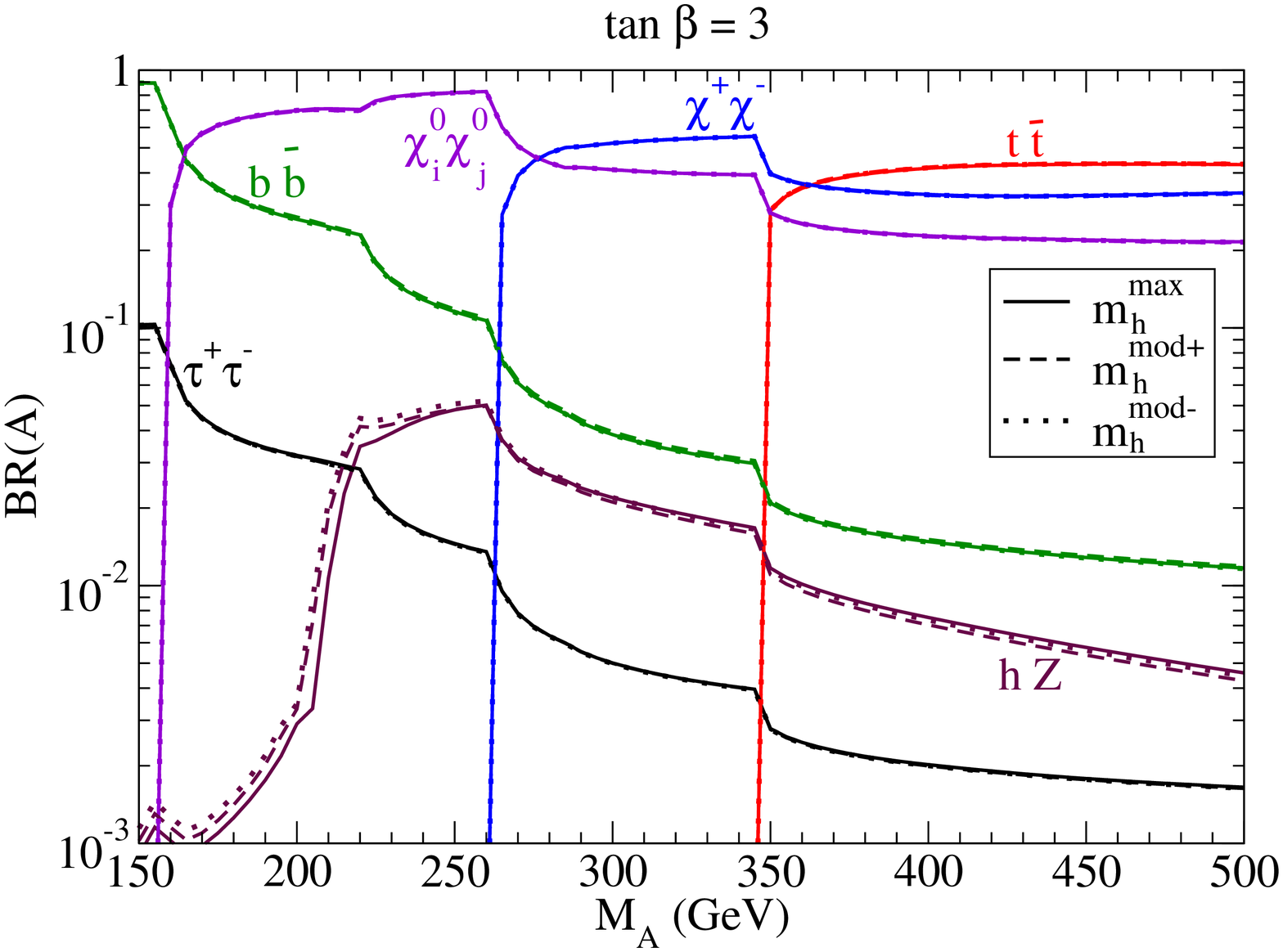}}
\subfigure[]{\includegraphics[width=0.3\textwidth,clip]{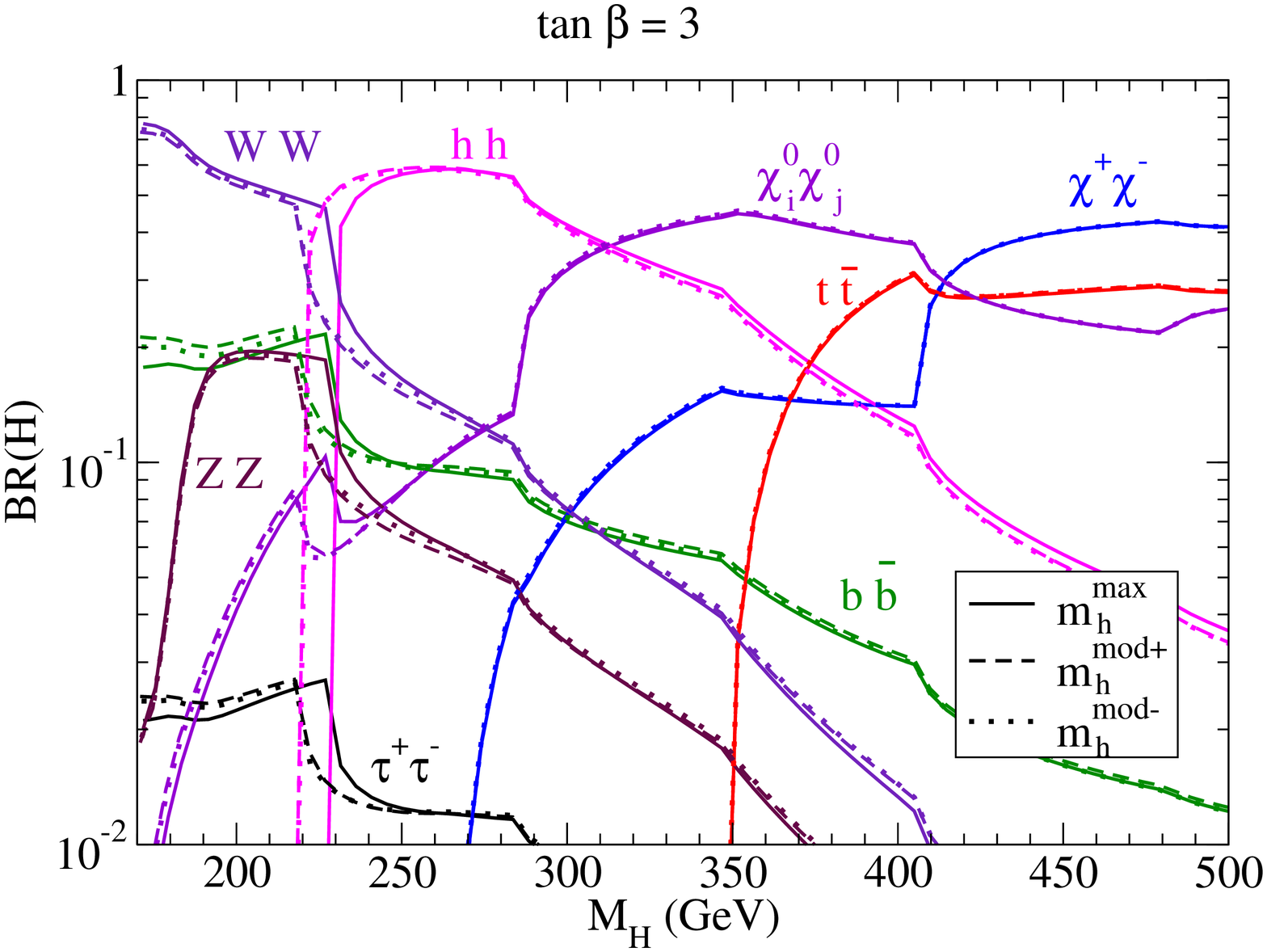}}
\caption{(a,d) Production rates at $\sqrt{S}=7$~TeV and (b,c,e,f) branching ratios of heavy neutral Higgs, $H$ and $A$, as a function of $M_{A,H}$ for (a-c) $\tan\beta=30$ and (d-f) $\tan\beta=3$.}
\label{ProdDec}
\end{figure*}

This correction to $M_h$ is maximized when $X_t=\sqrt{6}M_S$.  Given the difficulty of obtaining a large Higgs mass with small fine-tuning in the MSSM, this relation was used as a basis for the original $m_h^{max}$ benchmark point~\cite{Carena:2002qg}, with all other SUSY parameters set in the TeV range.  However, now that $M_h$ is known, we can see that $m_h^{max}$ actually produces too large a Higgs mass for much of the $M_A-\tan\beta$ plane (with TeV scale stops.)  With this in mind, additional new benchmarks have been recently proposed~\cite{Carena:2013qia}.  

In this study we extrapolate current LHC bounds on direct detection of the two heavy neutral Higgs bosons, $H$ and $A$, of the MSSM to 300 fb$^{-1}$ and 3000 fb$^{-1}$, with particular attention payed to $H/A\rightarrow\tau^+\tau^-$.  We focus on the benchmark points~\cite{Carena:2013qia}:

\begin{tabular}{clll}\\
$\bullet$&$m^{max}_h$: &$X^{\rm OS}_t=2M_{\rm SUSY}$ &  $X^{\overline{\text{MS}}}_t=\sqrt{6}M_{\rm SUSY}$\\\\
$\bullet$&$m^{mod+}_h$: &$X^{\rm OS}_t=1.5M_{\rm SUSY}$ &  $X^{\overline{\text{MS}}}_t=1.6M_{\rm SUSY}$\\\\
$\bullet$&$m^{mod-}_h$: &$X^{\rm OS}_t=-1.9M_{\rm SUSY}$ &  $X^{\overline{\text{MS}}}_t=-2.2M_{\rm SUSY}$,\\\\
\end{tabular}

\noindent
where $X^{\rm OS}_t$ has been calculated in the on-shell scheme and $X^{\overline{\text{MS}}}_t$ has been calculated in the $\overline{\text{MS}}$ scheme~\cite{Heinemeyer:1998yj,Carena:2000dp}.
The sfermion parameters are third generation squark masses $M_{\rm SUSY}=M_{\tilde{t}_L}=M_{\tilde{b}_L}=M_{\tilde{t}_R}=M_{\tilde{b}_R}=1$~TeV, third generation slepton mass $M_{\tilde{\ell}_3}=1$~TeV, third generation trilinear terms $A_t=A_b=A_\tau$, first and second generation masses $M_{\tilde{q}_{1,2}}=3M_{\tilde{\ell}_{1,2}}=1.5$~TeV, and first and second generation trilinear terms $A_{\tilde{f}_{1,2}}=0$.  The Higgsino and electroweak gaugino mass parameters are $\mu=M_2=200$~GeV with $M_1$ set by the GUT relation~$M_1=5/3\times \sin^2\theta_{W}/\cos^2\theta_W\times M_2$.  Finally, the gluino mass parameter $M_{\tilde{g}}=1.5$~TeV. This $m^{max}_h$ scenario differs slightly from the original~\cite{Carena:2002qg}.  While we focus on a few benchmark scenarios and direct detection of the heavy neutral Higgs bosons at the LHC, it is also possible to perform a more exhaustive scan over the phenomenological MSSM~\cite{Djouadi:1998di} parameter space and, in addition to direct constraints, apply indirect constraints such as flavor physics, dark matter, and measurements of the light Higgs branching ratios~\cite{Arbey:2013jla}.

The MSSM spectrum, branching ratios, and partial widths for this study have been obtained via FeynHiggs~\cite{Heinemeyer:1998yj}.  The production of heavy MSSM Higgs bosons occurs predominantly through gluon fusion via top and bottom quark loops and $b\bar{b}$ annihilation.  The production rates of the $H$ and $A$ are calculated by rescaling the SM production rate by the appropriate ratio of the MSSM and SM Higgs partial widths. The SM Higgs partial widths and gluon fusion production rate are taken from the LHC Higgs Cross Section Working Group~\cite{Dittmaier:2011ti}. The bottom quark annihilation the rate is calculated at NNLO in QCD with bbh@nnlo~\cite{Harlander:2003ai}.

In Fig.~\ref{ProdDec} we show branching ratios and LHC production cross sections of $H$ and $A$ for $\tan\beta=30$ and $\tan\beta=3$,  and the $m_h^{max}$ and $m_h^{mod\pm}$ scenarios.  The main features are:
\begin{itemize}
\item   At high $\tan\beta$, the coupling of the heavy Higgs bosons to down-type fermions is enhanced.    Hence, the bottom quark annihilation production mechanism dominates over the gluon fusion.  As $\tan\beta$ lowers, the coupling of $H/A$ to top quarks becomes more important and the top quark loop contribution to $gg$ fusion eventually dominates over bottom quark annihilation.  For $\tan\beta=3$, the cusp in $gg\rightarrow H/A$ at $M_{H/A}\approx 320$~GeV is due to an imaginary part present in the loop integral for $M_{H/A}>2m_t$.

\item For large $\tan\beta$ the decays to $\tau^+\tau^-$ and $b\bar{b}$ are dominant, again, because the heavy Higgs bosons couplings to down-type fermions are enhanced.  Since bottom quark final states are difficult to detect at the LHC due to large QCD backgrounds, for high $\tan\beta$ the main search channel for $H$ and $A$ is the $\tau^+\tau^-$ decay.  At low $\tan\beta$, there are many more decay channels with substantial branching ratios and a much richer phenomenology. 
\item As can be clearly seen, the production and decay rates for the $m_h^{max}$ and $m_h^{mod\pm}$ benchmarks are similar, in particular for lower $\tan\beta$ and $\tau^+\tau^-$ decays, which are our primary concern.  Hence, for the rest of the study we focus on the $m_h^{max}$ scenario.
\end{itemize}

\begin{figure*}[tb]
\centering
\subfigure[]{\includegraphics[width=0.3\textwidth,clip]{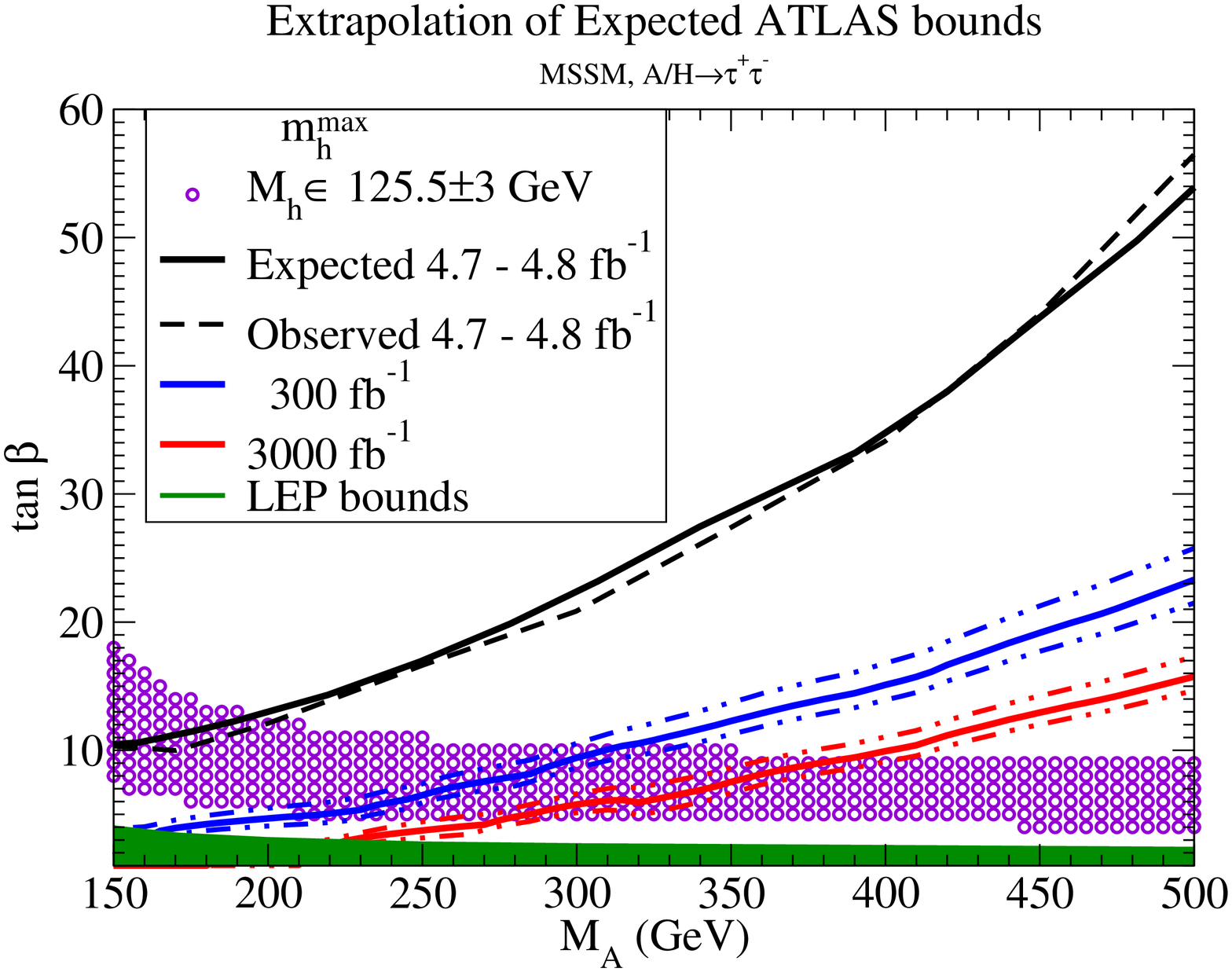}}
\subfigure[]{\includegraphics[width=0.3\textwidth,clip]{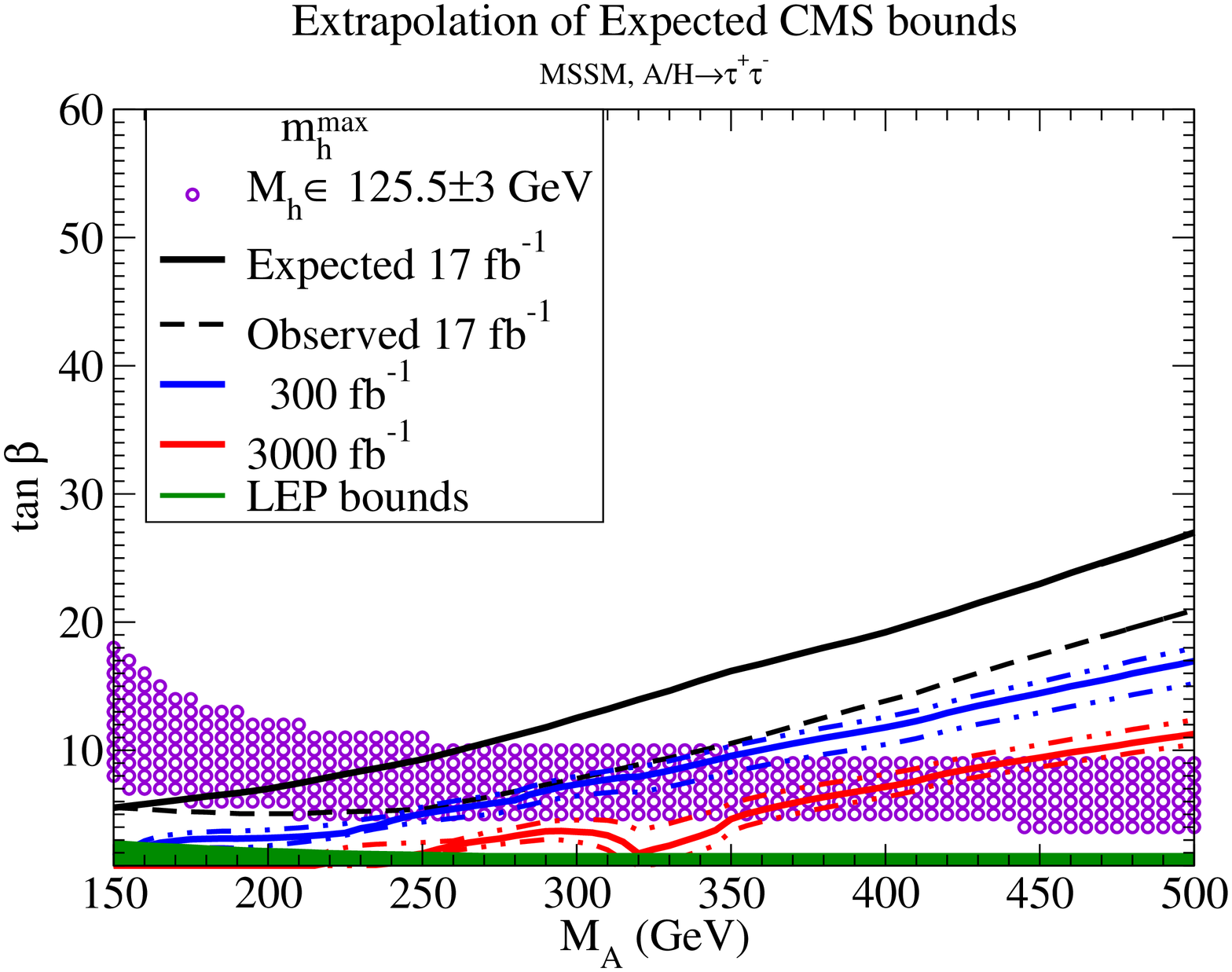}}
\subfigure[]{\includegraphics[width=0.3\textwidth,clip]{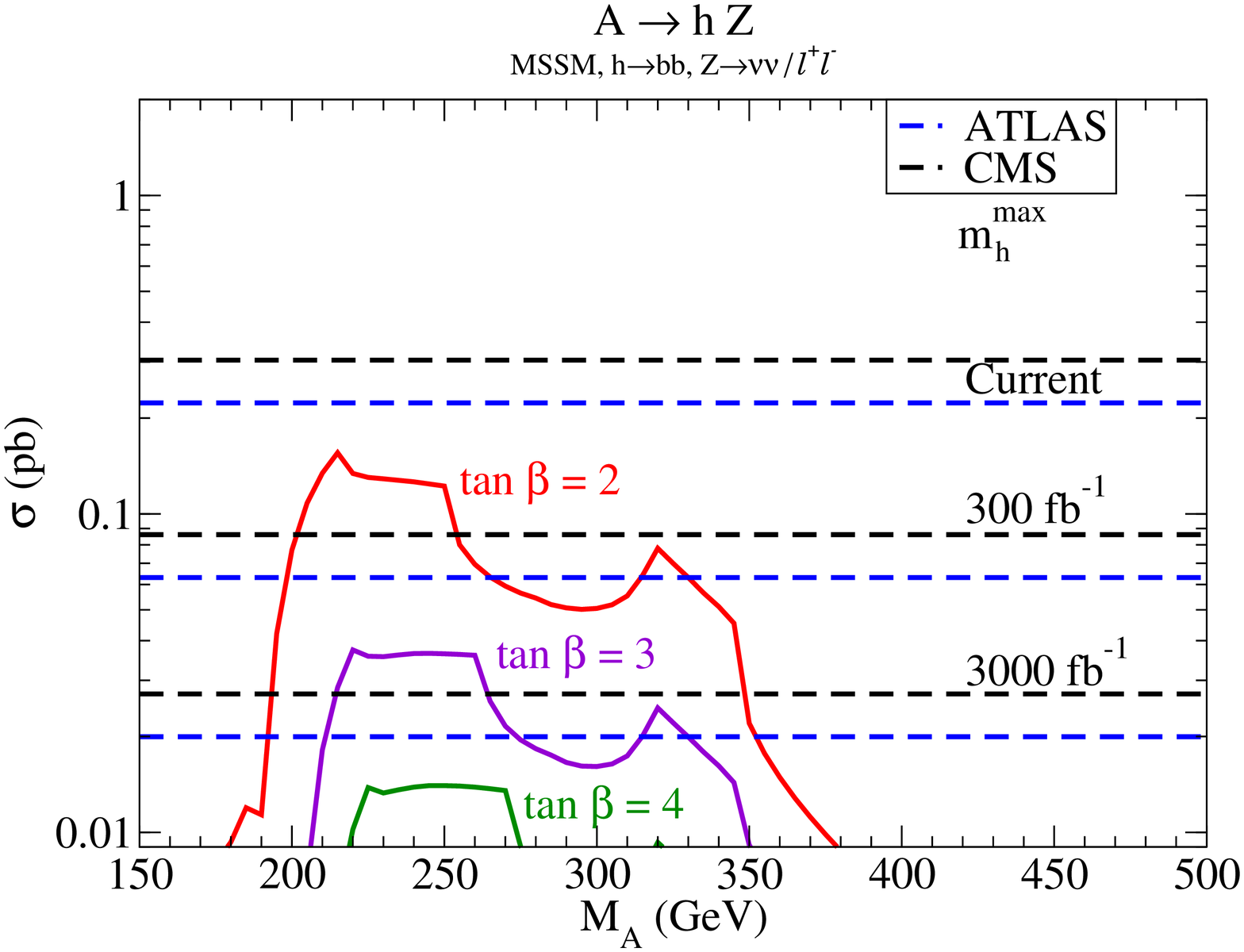}}\\
\subfigure[]{\includegraphics[width=0.3\textwidth,clip]{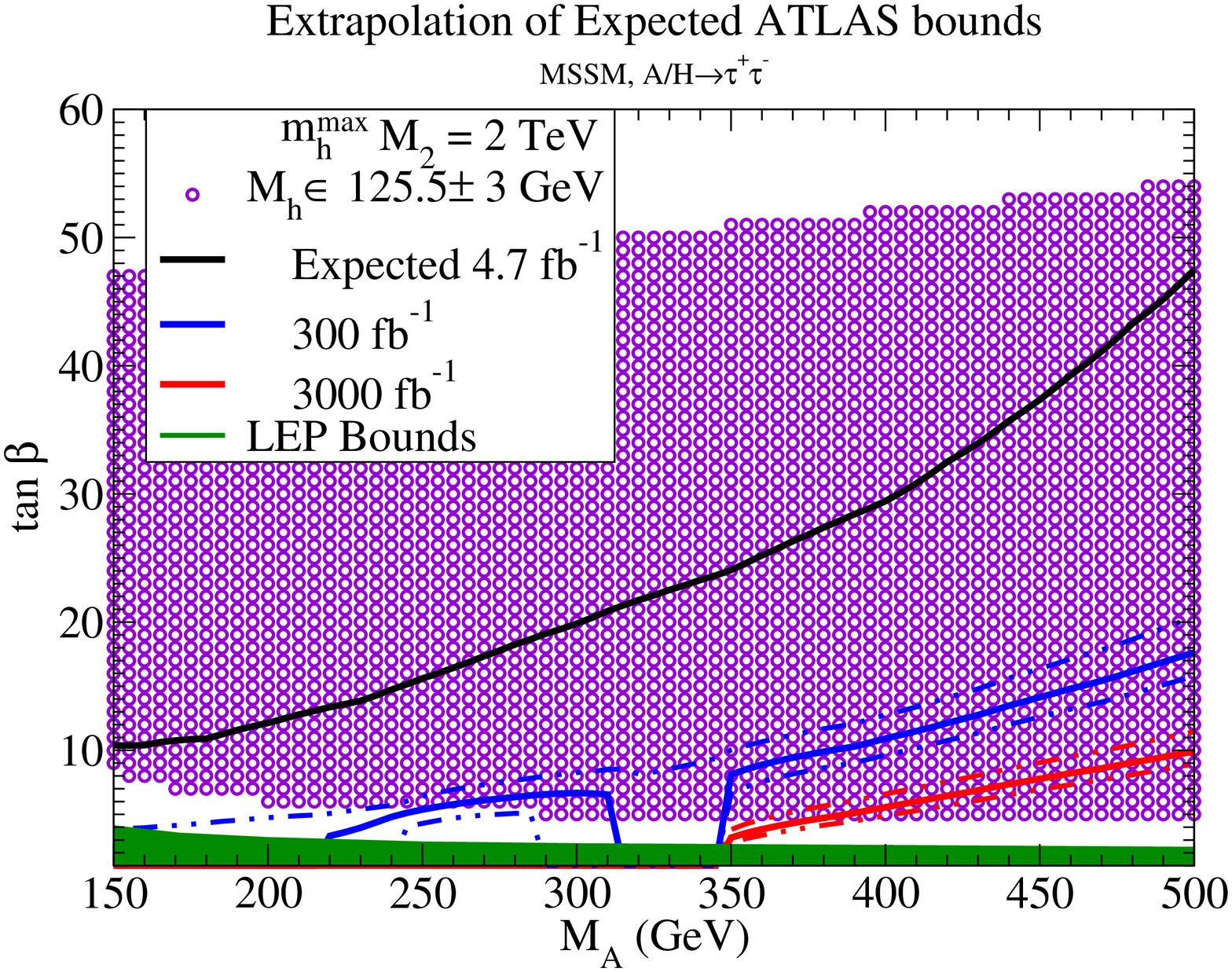}}
\subfigure[]{\includegraphics[width=0.3\textwidth,clip]{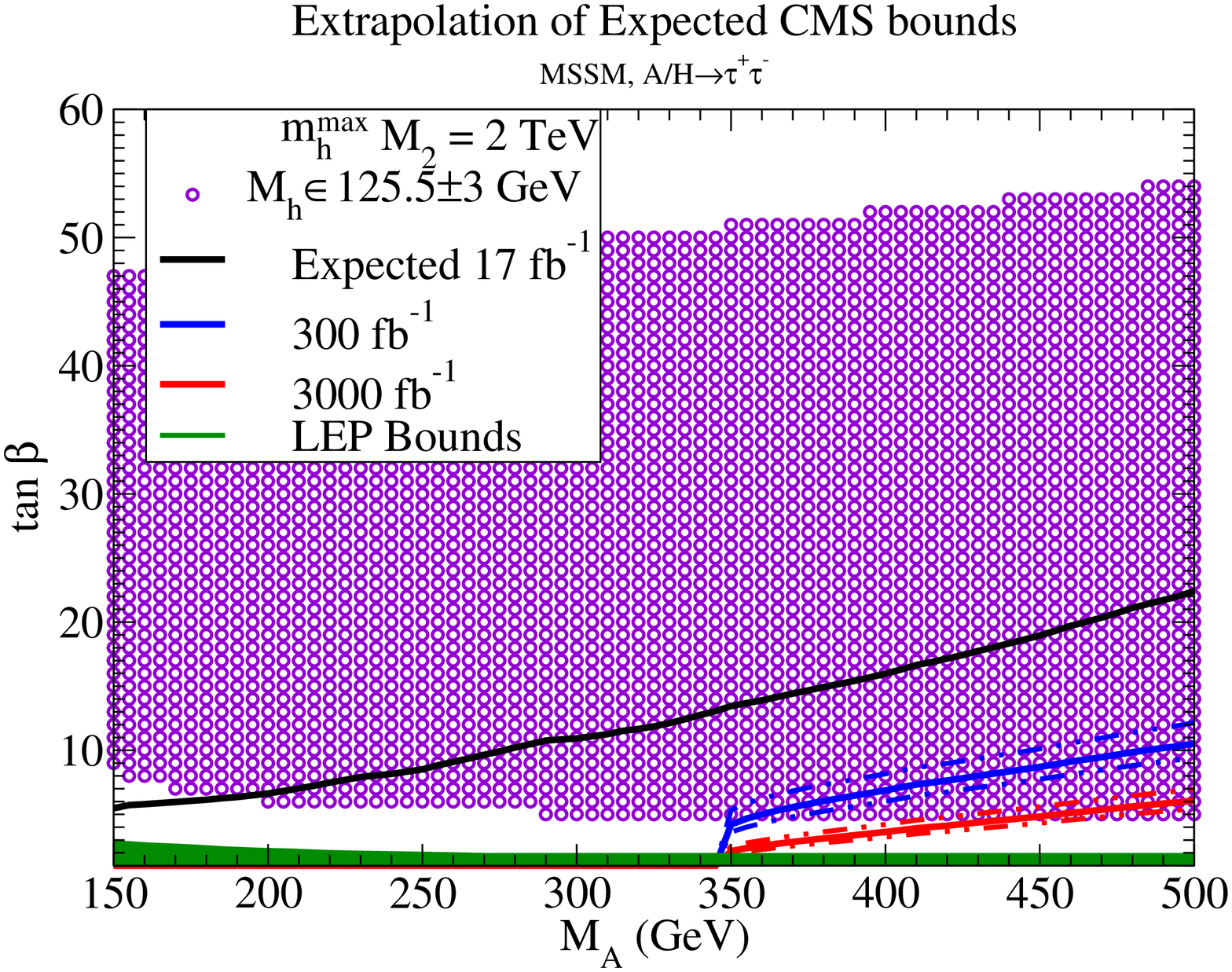}}
\subfigure[]{\includegraphics[width=0.3\textwidth,clip]{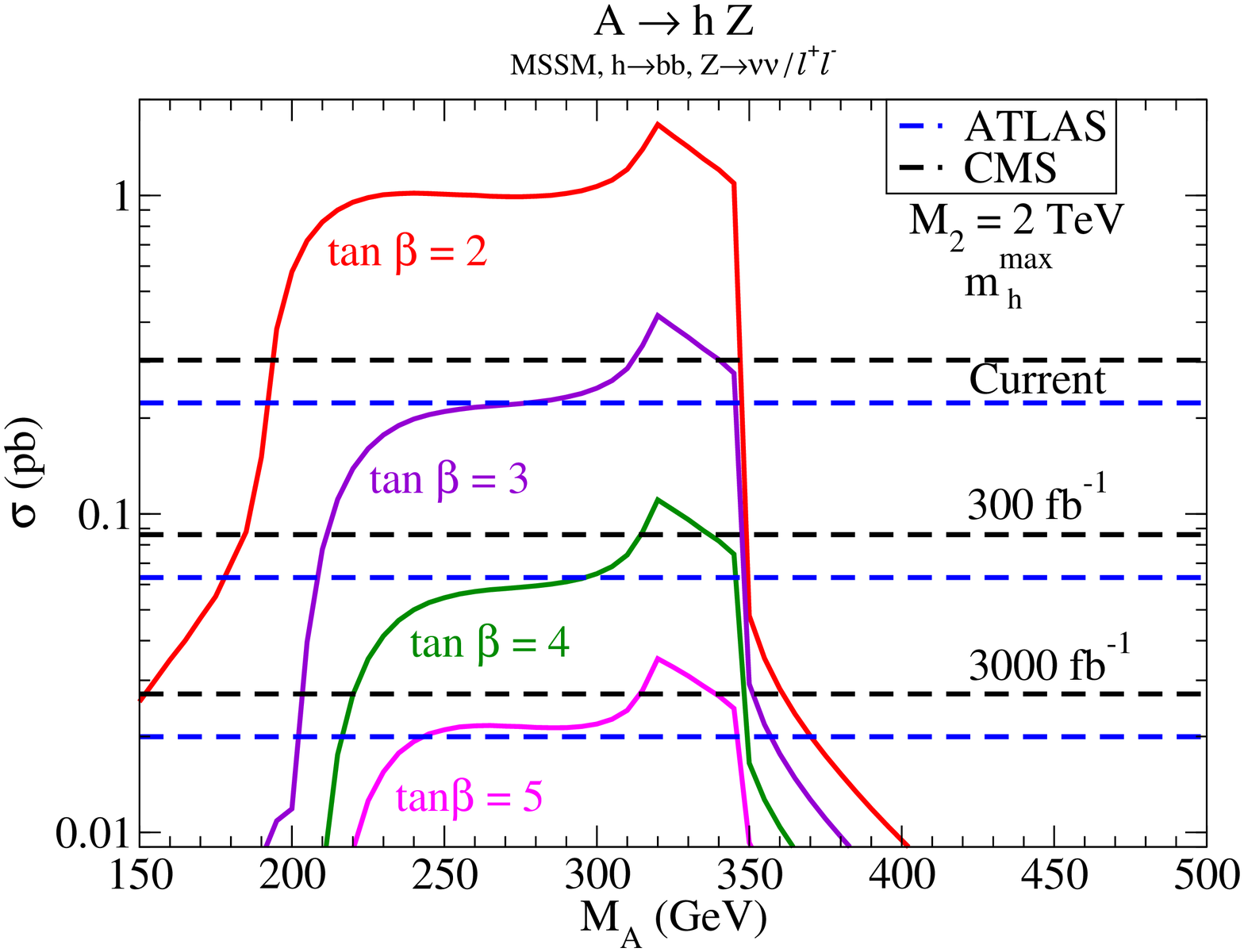}}
\caption{Expected exclusions from $H/A\rightarrow\tau^+\tau^-$ and $A\rightarrow hZ$.}
\label{bounds}
\end{figure*}

The heavy MSSM Higgs bosons have been searched for in the $H/A\rightarrow \tau^+\tau^-$ decay channel at ATLAS with 4.7~fb$^{-1}$ at 7 TeV~\cite{Aad:2012yfa} and at CMS with 4.9 fb$^{-1}$ at 7 TeV plus 12.1 fb$^{-1}$ at 8 TeV~\cite{CMS:gya}.  The published bounds in the $\tan\beta-M_A$ plane are for the original $m_h^{max}$ scenario~\cite{Carena:2002qg}, which differs from the new benchmark in the choice of the $M_{\tilde{g}}$.  We recast the bounds for the new $m_h^{max}$ scenario and extrapolate the expected bounds to 300 and 3000 fb$^{-1}$ by simply scaling the current expected cross section limits with the square root of luminosity.  Recasting the current bounds into the new $m_h^{max}$ scenario makes little change to the current bounds.

\begin{itemize}
\item The results of bounds in the $\tan\beta-M_A$ plane from the $\tau^+\tau^-$ extrapolation in the $m_h^{max}$ scenario are shown in Fig.~\ref{bounds}(a,b).  We show the up-to-date expected bounds (black solid), the up-to-date observed bounds (black dashed), and the expected bounds extrapolated to 300~fb$^{-1}$ (red) and 3000~fb$^{-1}$ (blue).  The area above and to the left of the curves are the projected exclusion.  The dash-dot-dot lines surrounding the extrapolation account for a theoretical error of $\pm 25\%$ in the calculation of $\sigma(H/A)\times \text{BR}(H/A\rightarrow \tau^+\tau^-)$~\cite{Baglio:2010ae,Djouadi:2013vqa}.  The dip in the 300 fb$^{-1}$ projections at $M_A\approx 320$~GeV is due to the cusp in the $gg\rightarrow H/A$ production cross section at low $\tan\beta$, as shown in Fig.~\ref{ProdDec}(d).  The purple dots indicate regions for which the $m_h^{max}$ scenario correctly reproduces the measured Higgs mass within 3~GeV.  A 3~GeV mass window is chosen to account for theoretical errors in the Higgs mass calculation.  The region in the $\tan\beta-M_A$ plane that correctly reproduces $M_h$ in the $m_h^{mod\pm}$ scenario fills above and most of the $m_h^{max}$ region.  As can be seen, with 300 fb$^{-1}$ (3000 fb$^{-1}$) ATLAS can exclude $M_A\lesssim 200-230$~GeV ($M_A\lesssim 260-290$ GeV) and CMS can exclude $M_A\lesssim 230-260$ GeV ($M_A\lesssim 290-360$ GeV) in the $H/A\rightarrow \tau^+\tau^-$ channel for the $m_h^{max}$ scenario with $M_{\rm SUSY}=1$~TeV.

\item By increasing the value of $M_2$, the neutralino and chargino masses increase making the decay of $H$ and $A$ into neutralinos and charginos kinematically impossible for much of the $M_A$ range presented here. Hence the branching ratio of $H,A$ into $\tau^+\tau^-$ increases.  In Figs.~\ref{bounds}(d) and (e) we show the ATLAS and CMS current and extrapolated expected bounds re-evaluated with $M_2=2$~TeV.   For $M_A\gtrsim 350$~GeV the $t\bar{t}$ decay is available and is dominant, decreasing the reach of the $\tau^+\tau^-$ bounds in this mass range. However, the bounds from $\tau^+\tau^-$ signals are still significantly increased.  In fact, in this channel with 3000 fb$^{-1}$ ATLAS can exclude $M_A<360-390$ GeV and CMS can exclude $M_A<410-460$~GeV for the $m_h^{max}$ scenario with $M_{\rm SUSY}=1$~TeV.

\item At lower $\tan\beta$, the branching ratio of $A\rightarrow hZ$ can be substantial.  Although there are no dedicated searches for $hZ$ resonances, the bounds on Higgs production in association with a vector boson, $Vh$, can be interpreted as a bound on the total $A\rightarrow hZ$ rate. Here we use the observed bounds on the SM rate as bounds on the total $A\rightarrow hZ$ rate. For $M_h=125$~GeV the current ATLAS~\cite{ATLAS} bound on the $Vh$ rate is 1.8 times the SM rate with 4.7~fb$^{-1}$ at 7 TeV plus 13 fb$^{-1}$ at 8 TeV, and the current CMS~\cite{CMSVH} bound is 1.9 times the SM rate with 5 fb$^{-1}$ at 7 TeV plus 19 fb$^{-1}$ at 8 TeV.  The current bounds are for $h\rightarrow b\bar{b}$ and leponic decays of the associated gauge boson.  The CMS analysis relies on boosted Higgs searches and so limits the range of $M_A$ for which the direct comparison is valid, but we include the result to make the point that the LHC is potentially sensitive to this channel for low $\tan\beta$.

Fig.~\ref{bounds}(c) shows the current bounds from CMS and ATLAS extrapolated to 300 and 3000 fb$^{-1}$ for various values of $\tan\beta$ in the $m_h^{max}$ scenario. The $Vh$ SM rate is taken from the LHC Higgs Cross Section Working Group~\cite{Dittmaier:2011ti}. The main difference between the CMS and ATLAS limits is that CMS include $W^\pm\rightarrow \tau\nu$ decays, increasing the $Vh$ SM cross section with which to compare. With 3000 fb$^{-1}$, the collaborations are sensitive to this channel for $\tan\beta\le3-4$.  Decoupling the SUSY decays with $M_2=2$~TeV increases this sensitivity to $\tan\beta\le 5$, as seen in Fig.~\ref{bounds}(d).  It should be noted that in this $\tan\beta$ region, the benchmark points do not correctly reproduce the light Higgs boson mass. This can be rectified by increasing $M_{\rm SUSY}$ as was recently studied in Ref.~\cite{Djouadi:2013vqa}. However, assuming that the other SUSY parameters are held fixed, we do not expect this to greatly effect the production or decay rates of $H$ and $A$,  and our projected bounds should be valid with this caveat.  The sensitivity of the LHC to this channel could be increased with a dedicated search.

\item As mentioned previously, at low $\tan\beta$ the phenomenology of the MSSM neutral heavy Higgs bosons becomes much richer.  For both $H$ and $A$ SUSY decays can become important and $H/A\rightarrow t\bar{t}$ has a substantial branching ratio when kinematically allowed.  Both ATLAS~\cite{ATLASttbar} and CMS~\cite{Chatrchyan:2012yca} have searched for heavy resonances decaying to top quark pairs, although only results for resonances with masses above $500-750$ GeV are publicly available. For $M_H$ between $2M_W$ and $2M_h$, $H$ dominantly decays to $W^+W^-$.  Similarly, $H$ has a non-negligible branching ratio $ZZ$ in the appropriate mass range. In this region, bounds on SM Higgs boson searches can be re-evaluated in terms of bounds on $H$ production. While kinematically allowed and below the $t\bar{t}$ and chargino thresholds, $H$ has a substantial branching ratio to pairs of light Higgs.  Although $hh$ resonances are not currently searched for, there may be some sensitivity to this channel~\cite{Djouadi:2013vqa}.

\end{itemize}

\acknowledgements{This work has been supported by the U.S. Department of Energy under grant
No.~DE-AC02-98CH10886.  I would like to thank S. Dawson for suggesting this project, and Chien-Yi Chen for useful discussions.}

%---------------------------------------------------------

\end{document}